\documentclass[twocolumn,aps]{revtex4}
\usepackage{amsmath}
\usepackage{amssymb}
\usepackage{graphicx}
\usepackage{dcolumn}
\usepackage{bm}
\usepackage{xcolor}
\usepackage{epstopdf}

\begin{document}

\preprint{Phys.Rev.B }

\title{Bulk and shear viscosities in multicomponent 2D electron system}
\author{A. D. Levin,$^1$ G. M. Gusev,$^1$ V. A. Chitta,$^1$ A. S. Jaroshevich,$^{2}$ and A. K. Bakarov $^{2,3}$}

\affiliation{$^1$Instituto de F\'{\i}sica da Universidade de S\~ao
Paulo, 135960-170, S\~ao Paulo, SP, Brazil}
\affiliation{$^2$Institute of Semiconductor Physics, Novosibirsk
630090, Russia}
\affiliation{$^3$Novosibirsk State University, Novosibirsk 630090,
Russia}

\date{\today}
\begin{abstract}
 We investigated magnetotransport in mesoscopic samples containing electrons from three different subbands in GaAs triple wells. At high temperatures, we observed positive magnetoresistance, which we attribute to the imbalance between different types of particles that are sensitive to bulk 
 viscosities. At low temperatures, we found negative magnetoresistance, attributed to shear viscosity. By analyzing the magnetoresistance data, we were able to determine both viscosities. Remarkably, the electronic bulk viscosity was significantly larger than the shear viscosity. Studying multicomponent electron systems in the hydrodynamic regime presents an intriguing opportunity to further explore the physics in systems with high bulk viscosity.

\end{abstract}
\maketitle
\section{Introduction}
The hydrodynamic description of the fermion-electron system diverges from kinetic theory and presents several intriguing predictions concerning electron transport in small-sized samples, especially in two dimensions. The pivotal concept that significantly advances our understanding of electronic transport phenomena is the notion that, under sufficiently strong electron-electron scattering, an effectively viscous hydrodynamics framework becomes applicable \cite{gurzhi, andreev, narozhny2}. Advances in materials science have enabled systematic investigations using exceptionally clean samples, facilitating the observation of a wide range of hydrodynamic effects.

These effects encompass resistance reduction with temperature (known as the Gurzhi effect) \cite{gurzhi, dejong, andreev, narozhny2, gusev1, gusev4, principi}, giant negative magnetoresistance \cite{alekseev1, narozhny1, dmitriev, narozhny2, gusev1, raichev2, gusev5, du, gusev6}, negative nonlocal resistance \cite{bandurin1, torre, pellegrino2, levin}, superballistic flow \cite{kumar, holder}, hydrodynamics with obstacles \cite{lucas, gusev7, gornyi, krebs}, photogenerated electron hole plasma  \cite{pusep, pusep2} and modifications to the Hall effect \cite{berdyugin, scaffidi, burmistrov, alekseev2, alekseev4, gusev2}. The recent progress overview in electronic hydrodynamics has been presented in the papers \cite{polini}-\cite{narozhny}.

In narrow channels, electron flow resembles Poiseuille flow of liquid in a pipe, where velocities near the walls approach zero. It has been established that the resistivity of a narrow strip follows the formula \cite{gurzhi}: $\rho=\frac{m}{ne^{2}}\eta\frac{12}{W^{2}}$,
 where $m$ denotes the effective mass, $n$ represents the density, $W$ signifies the strip width, and $\eta$ stands for shear viscosity; the shear viscosity can be derived using the Kubo formula \cite{bradlyn}. For a two-dimensional system, it is given by $\eta = \frac{1}{4}v_{F}^{2}\tau_{2,ee}$, where $v_{F}$ denotes the Fermi velocity and $\tau_{2,ee}$ represents the shear stress relaxation time due to electron-electron scattering, with the subscript "2" indicating that the viscosity coefficient is determined by the relaxation of the second harmonic of the distribution function \cite{dmitriev}.

References \cite{principi, burmistrov, burmistrov2} provides a coherent microscopic calculation of bulk viscosity for two-dimensional electrons in a sample featuring defects of small radius. The bulk or second viscosity, denoted as $\zeta$, characterizes the dissipation that occurs within a liquid when it experiences a uniform compression-like deformation. It has remained one of the controversial subjects of fluid dynamics \cite{sharma}. Understanding the volume viscosity is crucial for grasping various fluid phenomena, such as sound attenuation in multiatomic gases and  the propagation of shock waves \cite{sharma}. However, in a monatomic gas at low density and in an electron Fermi liquid, volume viscosity is zero \cite{khalatnikov}. Nevertheless, some common fluids exhibit bulk viscosities that are hundreds to thousands of times greater than their shear viscosities \cite{cramer}. Deriving bulk viscosity experimentally is complex.  To study bulk viscosity in Fermi liquid, several challenges must be addressed. First, it is essential to identify a system where the bulk viscosity is significantly enhanced and measurable. Second, it is necessary to determine how this effect can be easily extracted experimentally. Multicomponent systems offer a useful platform for exploring bulk viscosity.

Recent demonstrations have highlighted the significant role of bulk viscosity in the viscous flow of a two-component electron fluid \cite{alekseev5}. Positive saturating magnetoresistance  has been predicted for a two-component electron fluid in narrow samples. For example, in a double quantum well, intersubband scattering, which transforms one type of particle into another, can lead to an imbalance in flow depending on the bulk viscosity. The magnetic field induces a transition from regimes of independent flows of the two fluid components to a regime where imbalance occurs near the edge regions \cite{alekseev5}. It is worth noting that the theory for a two-component system can be readily extended to triple and multi-component electron systems.

Previous studies have explored hydrodynamic transport in two-component electron-hole systems within compensated semimetals \cite{alekseev6}. It has been shown that recombination effects near the edge are crucial, contributing to linear positive magnetoresistance. The interaction between shear viscosity and recombination effects in mesoscopic compensated semimetals has been explored in \cite{alekseev7}, but the impact of bulk viscosity was not examined.

In this study, we employed a triple-well system with high barrier heights. Transport measurements in this system reveal significantly different scattering times between the central well and the lateral wells. Electrons in the central well experience rare transitions to the lateral wells during scattering events, which is a necessary condition for the model proposed in \cite{alekseev5}. We compared these results with those from a triple-well system with low barriers, where the scattering times are nearly equal across all wells and we did not observe positive magnetoresistance induced by bulk viscosity. This is because the conditions of the theory are not met, as the transport properties in different subbands are not significantly different. Additionally, in our previous studies \cite{gusev6}, we did not observe positive magnetoresistance in a double-well system, as the necessary conditions were not fully met. By comparing the theoretical predictions from \cite{alekseev5} with the observed positive magnetoresistance at high temperature, we were able to derive the bulk viscosity. Notably, the bulk viscosity was found to be much larger than the shear viscosity.   

Additionally, we investigate negative magnetoresistance at lower temperature resulting from the magnetic field's influence on shear viscosity. We also determine the characteristic shear stress relaxation time of electrons, which is influenced by electron-electron scattering. 
\section{Experimental results and Discussions}
Our samples consist of symmetrically doped GaAs triple quantum wells (TQWs), separated by $Al_xGa_{1-x}As$ and $AlAs$ barriers. They feature a high total electron sheet density of $n_{s} \approx 9 \times 10^{11}  \text{cm}^{-2}$ and mobilities of $4.5 \times 10^{5}  \text{cm}^2/\text{Vs}$. The central well is approximately 220 $\AA$ wide, with both side wells having equal widths of 100 $\AA$. The barrier thickness $d_{b}$ is  20 $\AA$ for AlAs (wafer A) and for AlGaAs (wafer B).

The transport properties of these triple wells have been extensively investigated, encompassing phenomena such as magneto intersubband oscillations in low magnetic fields, microwave-induced oscillations, and the fractional quantum Hall effect \cite{wiedmann, wiedmann2, wiedmann3}. Detailed parameters of the quantum wells, including mobility, density, and structural characteristics, can be found in the supplementary materials \cite{suppl}.

\begin{figure}[ht]
\includegraphics[width=9cm]{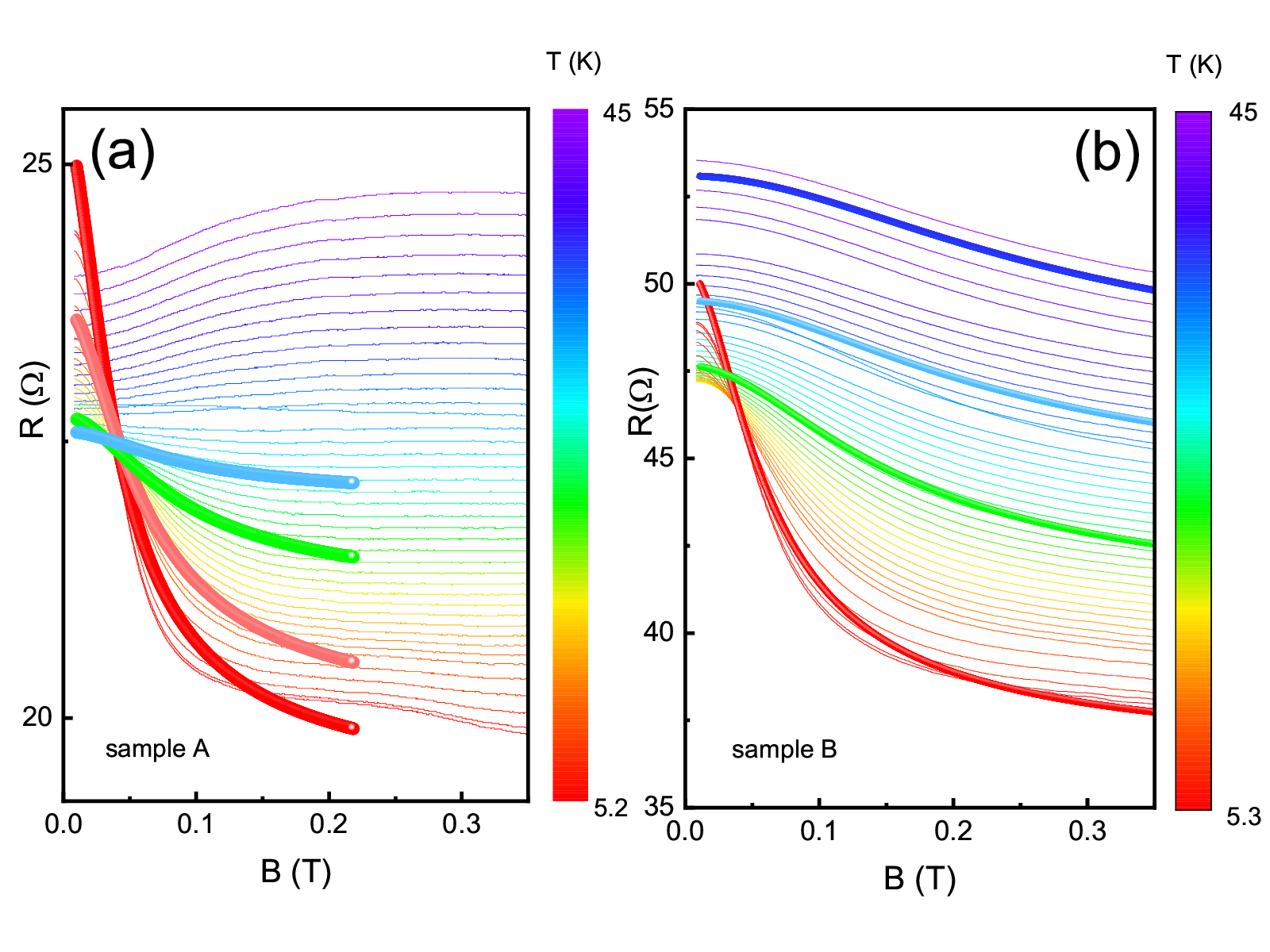}
\caption{\label{fig3}( Color online)
(a) Temperature-dependent magnetoresistivity of triple quatum wells (sample A). The circles (thick lines) are examples illustrating magnetoresistance calculated from Eq. (1) for different temperatures
T (K):5.3 (red), 11 (orange), 21 (green), 27 (cyan). (b) Temperature-dependent magnetoresistivity of triple quantum wells (sample B). The circles (thick lines) are examples illustrating magnetoresistance calculated from Eq. (1) for different temperatures
T (K):5.3 (red), 23 (green), 35 (cyan), 44 (blue). }
\end{figure}
\begin{figure}[ht]
\includegraphics[width=9cm]{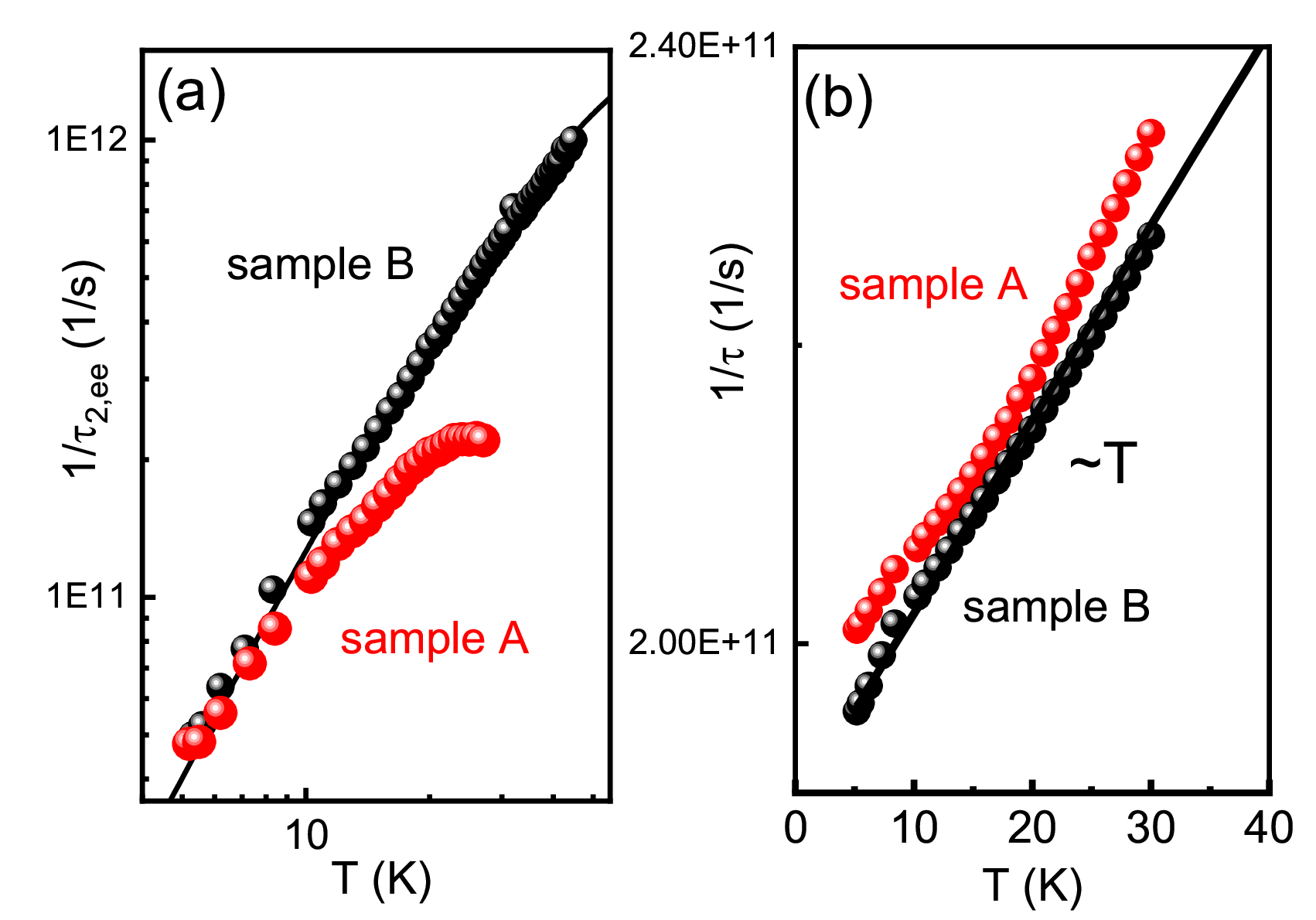}
\caption{\label{fig3}(Color online)
(a) Relaxation rate $1/\tau_{2,ee}$ as a function of temperature obtained by fitting the theory with experimental results for sample A (red circles) and sample B (black circles). Black line -theory. (b) Relaxation rate $1/\tau$ as a function of temperature obtained by fitting the theory with experimental data.}
\end{figure}
The average mean free path in macroscopic samples approaches  $l=(6-3.8) \mu$m, which exceeds the sample width $W$ \cite{suppl}. Hence, the hydrodynamic condition $W < l$ is satisfied. The mesoscopic sample is a Hall bar device featuring two current probes and seven voltage probes. The bar has a width  $W$  of 3.2 $\mu m$, and consists of three consecutive segments with different lengths, $L$, of 2.8  $\mu m$, 8.6 $\mu m$ and 32 $\mu m$. Details of the sample geometry, configuration, and measurement techniques are provided in \cite{suppl}. Measurements were conducted on two samples of each type of wafer (A and B), e.i. in 4 samples. We present the results for two samples, as the results for the other two samples are identical. 

Figure 1a depicts the evolution of resistance with magnetic field at various temperatures for sample A, which was fabricated using TQW with AlAs barriers. To improve hydrodynamic characteristics, current was applied between side probes, while voltage was measured across opposite side probes, utilizing an H-type geometry as described in \cite{gusev1, suppl}. It is noteworthy that the large negative magnetoresistance, characterized by a Lorentzian profile (i.e., $R(B)- R(0) < 0$), diminishes in magnitude and broadens as temperature rises. Furthermore, the resistivity at zero magnetic field exhibits a decrease with temperature within the range of $ 5 K < T < 25 K $ (known as the Gurzhi effect), followed by a subsequent increase. This observation aligns with earlier findings interpreted as distinct hallmarks of hydrodynamic behavior \cite{gusev1}. Figure 1b illustrates the pronounced negative magnetoresistance observed in sample B with $Al_xGa_{1-x}$As barriers. It is noteworthy that the behavior of Sample A contrasts distinctly with that of Sample B: above 25 K, the negative magnetoresistance diminishes and is replaced by significant positive magnetoresistance, saturating above 0.2 T. The resistance at zero field for Sample A is found to be two times smaller than that of Sample B, despite the fact that all macroscopic characteristics of the samples are almost identical.
\begin{figure}[ht]
\includegraphics[width=9cm]{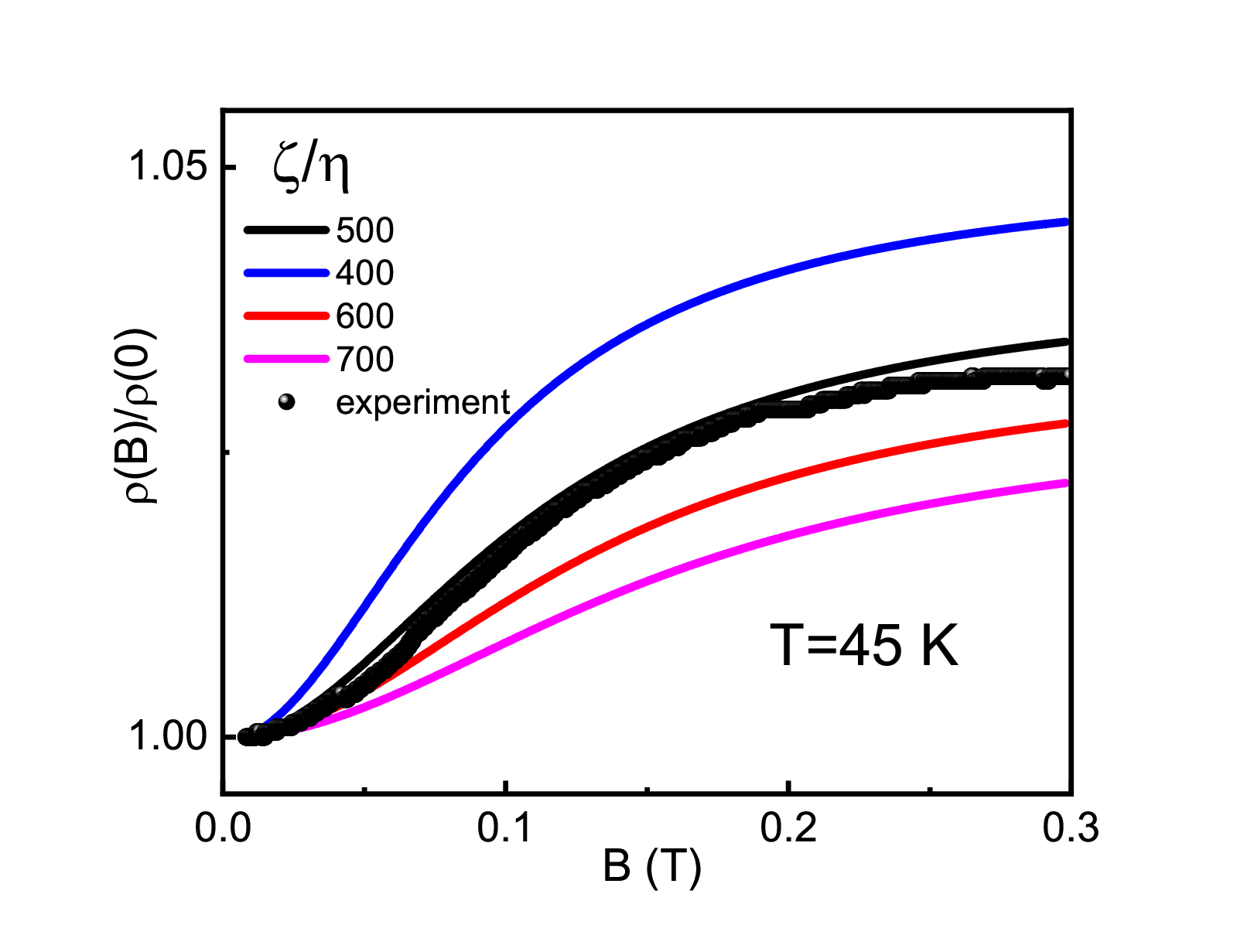}
\caption{\label{fig3}( Color online)
Dependence of the relative sample resistivity $\rho(B)/\rho(O)$ on magnetic field B for several values of bulk viscosity $\zeta$ (black circles), T=45K (sample A). Solid line - theoretical magnetoresistance calculated for different ratios of $\zeta/\eta$.}
\end{figure}
\begin{table}[ht]
\caption{\label{tab1} Fitting parameters of the electron system  for different configurations. Parameters are defined in the text.}
\begin{ruledtabular}
\begin{tabular}{lccccc}
&Sample &$1/\tau_{2,imp}$&$1/\tau_{0,imp}$ &  $B_{ph}$&$W^{*}$   \\
& & $(10^{11} 1/s)$ & $(10^{10} 1/s)$   &($10^{9} 1/sK$)  & $\mu m$  \\
\hline
&A& $1.1$  & $2$ &  1& 7.5\\ 
&B & $2.1$  & $3.0$  & 1.3& 3.5\\
\end{tabular}
\end{ruledtabular}
\end{table}
We attribute the negative magnetoresistance in both samples at low temperatures to hydrodynamic effects arising from shear electron viscosity. Conversely, the positive magnetoresistance for Sample A at high temperatures is attributed to transverse current imbalance governed by bulk viscosity, in accordance with predictions \cite{alekseev5}.

To qualitatively compare with experimental data, we utilize a model proposed in previous studies, originally developed for Poiseuille flow in the presence of a magnetic field in a single conductive layer \cite{alekseev1, alekseev5}. We have adapted this model for our specific configuration, which includes a common contact across all three layers.
Resistance can be expressed in the form: $R=\frac{L}{W}\rho_{total}=(\rho_{1}^{-1}+ \rho_{2}^{-1}+ \rho_{3}^{-1})^{-1}$,  and $\rho_{i}$ is given by equation below.
In its simplified form, the model characterizes resistivity through two distinct contributions. The first arises from ballistic effects or scattering at boundaries and defects, while the second is controlled by viscosity \cite{alekseev1}. In a more comprehensive formulation, the theory involves employing a viscosity tensor that depends on the magnetic field in order to derive the resistivity tensor for the i-th subband:
\begin{equation}
\rho_{i}(B)=\left(\frac{m}{e^{2}n_{i}}\right)\frac{1}{1-\tanh(\xi_i)/\xi_i}
\end{equation}
Here,  the dimensionless Gurzhi parameter $\xi=\xi_{0}\sqrt{1+(2l_{2}/r_c})^2$, $\xi_{0}=W^{*}/l_{G}$, where $l_{G}=\sqrt{l_{2}l}$ is the Gurhi lengh, $l_{2}=v_{F}\tau_{2}$, $l=v_{F}\tau$, $r_{c}=v_{F}/\omega_{c}$ is the cyclotron radius, $\omega_{c} = eB/mc$ represents the cyclotron frequency, $W^{*}$ is the effective sample width. The relaxation rate for shear viscosity is given by $1/\tau_{2}(T) = 1/\tau_{2,ee}(T) + 1/\tau_{2,imp}$.  The momentum relaxation rate is expressed as
$1/\tau(T) = 1/\tau_{0,ph}(T) + 1/\tau_{0,imp}$, where $1/\tau_{0,ph}=B_{ph}T$ represents the term associated with phonon scattering, and $1/\tau_{0,imp}$ denotes the scattering time due to static disorder (not related to the relaxation time of the second moment) \cite{alekseev1, alekseev}. Next, we fit the magnetoresistance curves and the resistivity  R(T) at zero magnetic field shown in Fig 1. For simplicity the fitting procedure employs three parameters: $\tau (T)$, $\tau_{2} (T)$ and the width of the sample $W$. In this case, we propose that the shear scattering time is nearly identical for different subbands; otherwise, fitting with numerous parameters would become excessively ambiguous.
We observe excellent agreement with Eq. (1) across a broad range of magnetic fields and temperatures for Sample B. However, the fitting of the data for Sample A is somewhat poorer, suggesting that additional mechanisms governing hydrodynamic properties may need to be considered. The parameters extracted from comparing the relaxation rates and theoretical equations are presented in Table 1. 

The total inelastic scattering rate arises from both inter-subband transitions and intra-subband processes, expressed as : $\left(1/\tau_{ee}\right)^{tot,i}= \left(1/\tau_{ee}\right)^{inter,i}+\left(1/\tau_{ee}\right)^{intra,i}$, where i=1,2,3 denotes the subband number. Electron-electron scattering is anticipated to be more intense due to enhanced screening effectiveness and a tripling of the phase space for intra-subband rates compared to the single-band scenario \cite{slutzky}. The inelastic scattering rate for  the intrasubband processes is given by:
$\left(\hbar/\tau_{ee}\right)^{intra,i}= -A_{1}(kT)^2/E_{F}+A_{2}[(kT)^2/E_{F}][ln\left(4E_{F}/kT\right)]$.

And for intersubband scattering:$\left(\hbar/\tau_{ee}\right)^{inter,i}= -B_{1}(kT)^2/E_{F}
+B_{2}[(kT)^2/E_{F}][ln\left(4E_{F}/\Delta\right)
+B_{3}[(kT)^2/E_{F}][ln\left(\Delta/kT\right)]$.
Everywhere $A_{j}$ and $B_{j}$ are positive numerical constants of order unity. The relaxation rate $1/\tau_{2,imp}(T)$, arising from processes relaxing the second harmonic of the distribution function, includes scattering by static defects contributing to viscosity. Conversely, $1/\tau_{2,ee}(T)$ corresponds to the relaxation of shear viscosity due to electron-electron scattering \cite{alekseev, alekseev1}. Figure 2a demonstrates the temperature dependence of the relaxation rate $1/\tau_{2,ee}(T)$ for both samples. For comparison with theory, we simplified the situation by assuming equal parameters for each subband and $1/\tau_{2,ee}(T) \approx  1/\tau_{ee}(T)$.  The solid line represents the theoretical comparison with parameters. Figure 1(a) shows the theoretical predictions with parameters  $A_{1} = B_{1} = 0.35$, and    $A_{2} = B_{2} = B_{3} = 0.26$. In sample A, we observe deviations in the scattering rate at high temperatures, suggesting that additional mechanisms are influencing the hydrodynamic flow in this sample.

We obtain $1/\tau_{1}\approx 1/\tau_{2}\approx 1/\tau_{3}\approx 1/\tau $ for sample B and  $1/\tau_{1}\approx 1/\tau_{2}\approx 1/\tau $, $1/\tau_{3}\approx 1/3\tau $ for sample A. Note that this ratio can vary and does not affect the shear relaxation time, which depends on  shape of the magnetoresistance. However, for other ratios, we obtain a coefficient $B_{ph}$ responsible for electron-phonon scattering, much higher than the value of $1\times10^9 1/sK$.  For example, if we assume that all subbands in sample A have a similar $1/\tau$ (as we proposed for sample B), the coefficient $B_{ph}$ is half the value reported in the literature or what we found in sample B. It is natural to assume that the penetration through the barriers depends exponentially on the height of the barrier. In the sample with low barriers, the wave functions for all subbands are spread across all three wells, so the scattering rate is expected to be nearly equal for all subbands. However, in sample A, with AlAs barriers, the wave functions are mostly localized in either the central or lateral wells, leading to different scattering rates. It demonstrates that electrons from different subbands in sample A exhibit nearly independent dynamics, with infrequent transformations into each other during scattering events. This characteristic is a necessary requirement for the model proposed in \cite{alekseev5}, which predicts hydrodynamic-induced positive magnetoresistance at high temperatures, as discussed below. Figure 2b illustrates the dependence of the rate $1/\tau$ on temperature. It shows a linear trend, consistent with theoretical expectations. The value of $B_{ph}$ agrees with previously calculated parameters that characterize the electron-phonon coupling in GaAs systems \cite{gusev1, levin}.

It is noteworthy that we also derive  the effective sample width  $W^{*}$ and obtain reasonable agreement for sample B: $W^{*}\approx3.5 \mu m$, which is slightly higher than geometric width  $W\approx 3 \mu m$. However, we observe a significantly different value for Sample A ( see Table 1), which further suggests that additional mechanisms are influencing the hydrodynamic flow in this sample.

To describe the positive magnetoresistivity at high temperatures for Sample A ( fig.1b), we utilize the model proposed in \cite{alekseev5}. This model examines a two-component electron liquid, where electron scattering and transitions between components can induce imbalances in flows and concentrations. This imbalance is particularly sensitive to bulk viscosity. Under the influence of a magnetic field, the system transitions from independent, uniform Ohmic flows of two carrier types to flows involving recombinations of these carriers, resulting in positive magnetoresistance. Note that the magnetic dependence of shear viscosity in this case is weak and negligible, as observed at elevated temperatures $T > 45 K$, where the shear relaxation rate is high. For a pure hydrodinamic case, the conductivity in the presence of the magnetic field is given by:
\begin{widetext}
\begin{equation}
\sigma_{hydr}(T,B)=\frac{e^2 n_{tot}}{m^*} \frac{1}{\left(\frac{n_{1}^{*}}{n_{tot}} \eta_1+\frac{n_{2}}{n_{tot}} \eta_2\right)}\left\{\frac{W^2}{12}+\frac{2 \zeta \lambda}{\omega_c^2} \frac{n_{1}^{*} n_{2} (\eta_1-\eta_2)^{2}}{n_{tot}\left(n_{1}^{*} \eta_1+n_{2} \eta_2\right)} \Re e\left[\sqrt{i} \tanh \left(\sqrt{i} \frac{\lambda W}{2}\right)\right]\right\} .
\end{equation}
\end{widetext}
Here, $\lambda=\sqrt{\omega_c} \sqrt[4]{\frac{a_s}{\zeta \eta}}$, where $1/\lambda$ represents the characteristic length that defines the widths of the near-edge regions, $\eta=\eta_{1}+\eta_{2}$. These regions are crucial for intense diffusion and particle type transformations, which in turn define the bulk viscosity $\zeta$ and facilitate the diffuse transport of momentum x-components in the y-direction perpendicular to  the channel due to shear viscosity effects in the fluid. The density $n_{tot}$ is the total carrier concentration, $\omega_c=eB/mc$ is the cyclotron frequency, $n_{i}$ and $\eta_{i}$ are corresponding density and viscosity of ith subbands. For simplicity, we propose that both lateral wells have similar density and viscosity, thus $n_{1}^{*}=2n_1$ and $\eta_{1} \approx \eta_{3} = \frac{1}{4}v_{F,1}^{2}\tau_{2,ee}, \quad \eta_{2} = \frac{1}{4}v_{F,2}^{2}\tau_{2,ee}$,
where $v_{F,i}$ denotes the Fermi velocity dependent on density, while $\tau_{2,ee}$ remains independent of the subband index.  At low magnetic field magnetic field, Equation 3 can be expressed as: $\sigma_{hydr}(T,0)=\frac{e^{2}}{m} \frac{W^2}{12}\left(\frac{n_{1}^{*}}{\eta_1}+\frac{n_{2}}{\eta_2}\right)$. This equation describes the Poiseuille flow of a uniform two-component fluid, incorporating contributions from both components' shear viscosities. In a strong magnetic field, Eq 3 transforms to $\sigma_{hydr}(T,\infty) = \frac{e^2 n_{tot}}{m^{*}} \frac{1}{\left(\frac{n_{1}^{*}}{n_{tot}} \eta_1 + \frac{n_{2}}{n_{tot}} \eta_2\right)} \frac{W^2}{12}$. This value corresponds to the sum of two independent Poiseuille flows involving different types of particles. The transition between these regimes corresponding to positive magnetoresistance (negative magnetoconductance) is a smooth crossover that occurs at magnetic field $B^{*}=mc \omega^{*}/e $ with the characteristic borderline cyclotron frequency: $\omega^{*}=\sqrt{\frac{\zeta \eta}{a_s}} \frac{1}{W^2}$, where $a_s=\frac{\eta}{n_{tot}}\left(\frac{n_{2}}{\eta_1}+\frac{n_{1}^{*}}{\eta_2}\right)$.  We expressed the total conductivity in the form $\sigma_{total}=\sum_{i=1}^{3}e^2 n_{i}\tau_i/m+\sigma_{hydr}$.  Figure 3 shows a comparison of the relative magnetoresistivity with the theoretical model \cite{alekseev5}, using parameters derived from the analysis of the negative magnetoresistance described in the previous sections (Fig.1). We approximate the temperature dependence of relaxation times  $\tau_{2,ee}$ and $\tau$ at high temperatures, using viscosity $\zeta$ as a fitting parameter. We find good agreement for the ratio $\zeta/\eta=500$, corresponding to $\zeta\approx 17 m^2/s$. Additionally, we determine $\eta_{1}=0.026 m^2/s$ and $\eta_{2}=0.0134 m^2/s$. These shear viscosity values are consistent with those previously reported for similar temperatures \cite{gusev1, levin}. It is well-known that the Boltzmann kinetic equation predicts a zero value for bulk viscosity in a monoatomic gas \cite{khalatnikov}. Estimating bulk viscosity in interacting Fermi gases has been addressed \cite{holten}, but extracting it from experiments remains an experimental challenge \cite{sharma}. In multi-component molecular liquids, a significant bulk
viscosity arises due to relatively slow, reversible chemical reactions between the liquid’s components \cite{cramer}.
\section{Conclusion}
In conclusion, we investigated hydrodynamic magnetotransport in triple quantum wells. In addition to observing negative magnetoresistance, we also noted positive magnetoresistance, attributed to an imbalance near the edges and governed by bulk viscosity. Both bulk and shear viscosities were deduced from our analysis of these magnetoresistances. Studying magnetoresistance in multicomponent systems paves the way for investigating large bulk viscosity, a parameter challenging to determine through other experiments.

We thank P. S. Alekseev  for helpful discussions. This work is supported by FAPESP (São Paulo Research Foundation) Grants No. 2019/16736-2, No. 2021/12470- 8, No. 2024/06755-8 and CNPq (National Council for Scientific and Technological Development).

\clearpage
\appendix
\section*{Supplemental Material: Bulk and shear viscosities in multicomponent 2D electron system}
\addcontentsline{toc}{section}{Supplemental Material:Bulk and shear viscosities in multicomponent 2D electron system}

In this supplemental material, we present detailed information regarding the sample description and transport measurements conducted in triple quantum well bars. Additionally, we discuss the hydrodynamic conditions in a mesoscopic triple well.

\maketitle
\section{Experimental details}
We studied triple quantum wells with $AlAs$ (Sampe A) and  $Al_{x}Ga_{1-x}As$  (Sample B) barriers.  Sample A consist of symmetrically doped GaAs triple quantum wells (TQWs), separated by $AlAs$ barriers. These layers exhibit a high total electron sheet density of $n_{s} =9.2 \times 10^{11}$ cm$^{-2}$ and avarage mobilities of $5 \times 10^{5}$ cm$^{2}$/V s. The central well width is approximately $220$ Å, with both side wells having equal widths of $100$ Å. The barrier thickness $d_b$ is $20$ Å.

Sample B is composed of symmetrically doped GaAs TQWs separated by $ Al_{x}Ga_{1-x}As$ barriers, with a total electron sheet density of $n_{s} = 8.5 \times 10^{11}$ cm$^{-2}$ and average mobilities of $4 \times 10^{5}$ cm$^{2}$/V s. The central well width is about $220$ Å, and both side wells have equal widths of $100$ Å. The barrier thickness $d_b$ is $20$ Å. It's important to note that the energy of AlAs barriers is significantly higher than that of $Al_{x}Ga_{1-x}As$ barriers. 
\begin{figure}[ht]
    \centering
    \includegraphics[width=8cm]{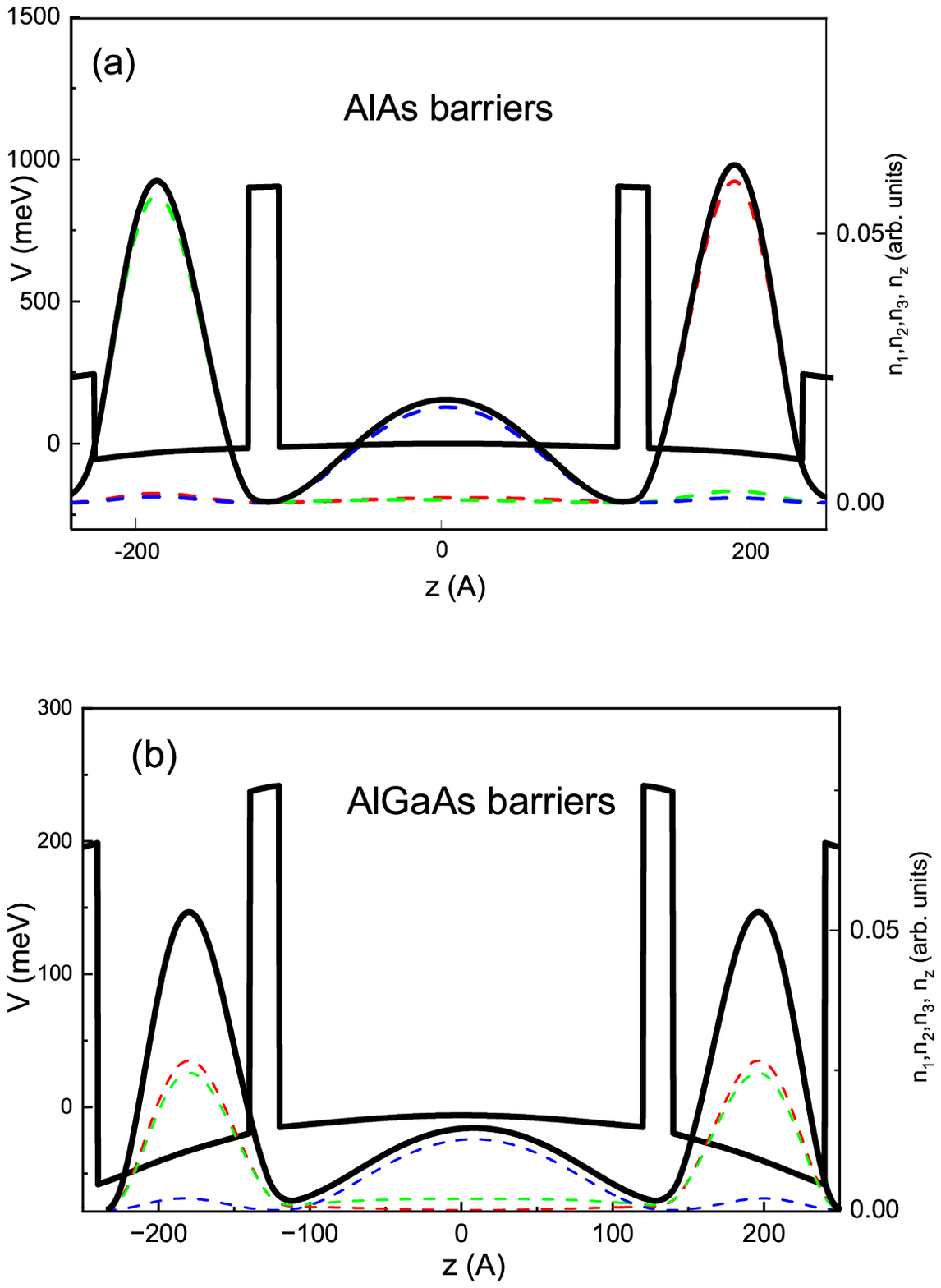}
    \caption{Profiles of a triple quantum well for $AlAs$ (a) and $Al_{x}Ga_{1-x}As$ barriers (b). The dashed lines indicate the profiles of the subband densities, while the solid lines represent the profile of the total density.}
 
\end{figure}
\begin{figure}[ht]
\includegraphics[width=8cm]{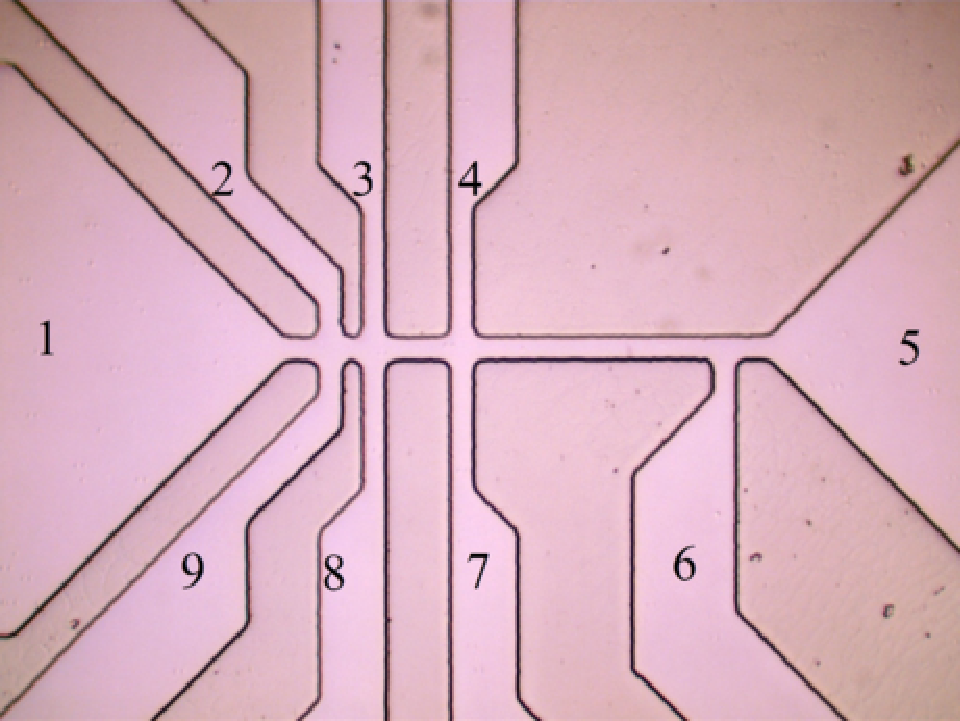}
\caption{\label{fig3}(Color online)
Image of the mesoscopic Hall bar device.}
\end{figure}
\begin{table*}[ht]
\begin{ruledtabular}
\begin{tabular}{lccccccccc}
TQW & $d_{l,r}$ & $d_{c}$ & $d_{b}$ & $N_{s,1}$ & $N_{s,2}$ & $N_{s,3}$ &$\Delta_{1,2}$ &$\Delta_{1,3}$ &$\Delta_{2,3}$ \\
    & (\AA)     & (\AA)   & (\AA)   & ($10^{11}$ cm$^{2}$) & ($10^{11}$ cm$^{2}$) & ($10^{11}$ cm$^{2}$) & (meV) & (meV)& (meV)\\
\hline
A & 100 & 220 & 20 (AlAs) & 3.2 & 3.2 & 2.9 &0.23& 2& 1.7 \\
B   & 100 & 220 & 20   & 3.6 & 3.2 & 2.2 & 1& 2.4& 3.4\\
\end{tabular}
\end{ruledtabular}
\caption{\label{tab1} Parameters of the triple quantum wells, $d_{l,r}$,$d_{c}$ is the width of the left(right) and the central wells,  $d_{b}$ is the barrier width, 
$N_{s}^{1,2,3}$ is the density for three occupied subbands (1,2,3),determined from Fourier analysis of magnetoresistance}
\label{tab:example}
\end{table*}
To populate the central well effectively, its width was increased. The estimated electron density in the central well is lower than that in the side wells. 

In TQWs with transparent barriers, the energies of quantization and the corresponding single-electron wave functions can be estimated using the tight-binding model. For a symmetric structure, the energy levels can be expressed as follows:
\begin{gather}\label{eq2} 
E_{1,3} = \frac{\varepsilon_c + \varepsilon_s}{2} \mp \sqrt{\frac{(\varepsilon_c - \varepsilon_s)^2}{4} + 2 t^2}, 
E_2 = \varepsilon_s 
\end{gather}
where $\varepsilon_s$ and $\varepsilon_c$ represent the energies of electrons in the side and central wells respectively, in the absence of tunneling, and $t$ denotes the tunneling amplitude. Given that $\varepsilon_c > \varepsilon_s$ and $t > 0$, the subband sequence follows as $E_1 < E_2 < E_3$.
The corresponding eigenstates are expressed using single-well orbitals as $\psi_j(z) = \sum_i \chi_{ij} F_i(z)$. $F_{i}(z)$ in the basis of single-well orbitals $F_i(Z)$, $i = 1 , 2 , 3$ numbers the left, central, and right well, respectively. The matrix $\chi_{ij}$ is provided by
$\chi_{i j}=\left(\begin{array}{ccc}
C_1 t /\left(\varepsilon_s-\varepsilon_1\right) & 1 / \sqrt{2} & C_3 t /\left(\varepsilon_s-\varepsilon_3\right) \\
C_1 & 0 & C_3 \\
C_1 t /\left(\varepsilon_s-\varepsilon_1\right) & -1 / \sqrt{2} & C_3 t /\left(\varepsilon_s-\varepsilon_3\right)
\end{array}\right)$

The parameters $C_{1,3}$ are defined as $\left( 1 + \frac{2t^2}{(\epsilon_s - \epsilon_{1,3})^2} \right)^{-1/2}$. 

The parameters of the tight-binding model can be extracted from experimentally determined subband gaps. Setting $\epsilon_s$ as the reference energy, we derive $\epsilon_c = \Delta_{23} - \Delta_{12}$ and $t = \sqrt{\frac{\Delta_{23} \Delta_{12}}{2}}$.

In Sample A with $AlAs$ barriers, quantum wells were populated with electrons of nearly equal density. All observation aligns with experimental findings confirmed by Shubnikov de Haas oscillations in a weak magnetic field \cite{wiedmann, wiedmann2, wiedmann3}.

For our sample B, we found $\epsilon_c = 2.5$ meV and $2t = 3.35$ meV. With a total density $n_s \approx 9 \times 10^{11}$ cm$^{-2}$, the subband densities are $n_1 = 3.62 \times 10^{11}$ cm$^{-2}$, $n_2 = 3.23 \times 10^{11}$ cm$^{-2}$, and $n_3 = 2.14 \times 10^{11}$ cm$^{-2}$. The electron density in each side well is $n_{\text{side}} = \sum_j \chi_{1j}^2 n_j = \sum_j \chi_{3j}^2 n_j = 3.23 \times 10^{11}$ cm$^{-2}$, and the electron density in the central well is $n_{\text{cent}} = \sum_j \chi_{2j}^2 n_j = 2.53 \times 10^{11}$ cm$^{-2}$.

In addition, we performed self-consistent calculations of the potential profiles and concentrations, as shown in Figs 4 a and b. Our results are consistent with a simplified tight-binding model.
The layers are connected with ohmic contacts, and the corresponding parameters are summarized in Table \ref{tab1}. 

The mesoscopic sample is a Hall bar device featuring two current probes and seven voltage probes. The bar has a width  $w$  of 3-3.2 $\mu m$, and consists of three consecutive segments with different lengths, $L$, of 2.8  $\mu m$, 8.6 $\mu m$ and 32 $\mu m$. Figure 2 shows the image of the mesoscopic Hall bar device.

For the measurements, we utilized a VTI cryostat and employed a conventional lock-in technique. This technique allowed us to measure the longitudinal resistance. To avoid overheating effects, we applied an alternating current (ac) of 0.1–1 $\mu A$ through the sample, which is considered sufficiently
low. The current I flows between contacts 1 and 5, and the voltage V was measured between probes 2 and 3, R = $R_{1,5}^{2,3} = V_{2,3}/I_{1,5}$ (Fig. 5). In  H-type geometry, the current I flows between contacts 9 and 8, and the voltage V was measured between probes 2 and 3, R = $R_{9,8}^{2,3} = V_{2,3}/I_{9,8}$. Furthermore, we compared our findings with the transport properties of two-dimensional (2D) electrons in a larger-scale sample. The mean free path of electrons
in the macroscopic sample is 3.8 $\mu m$ at T = 4.2 K, which exceeds the width of the sample. 
\section{Hydrodynamic conditions}
\begin{figure}[ht]
\includegraphics[width=10cm]{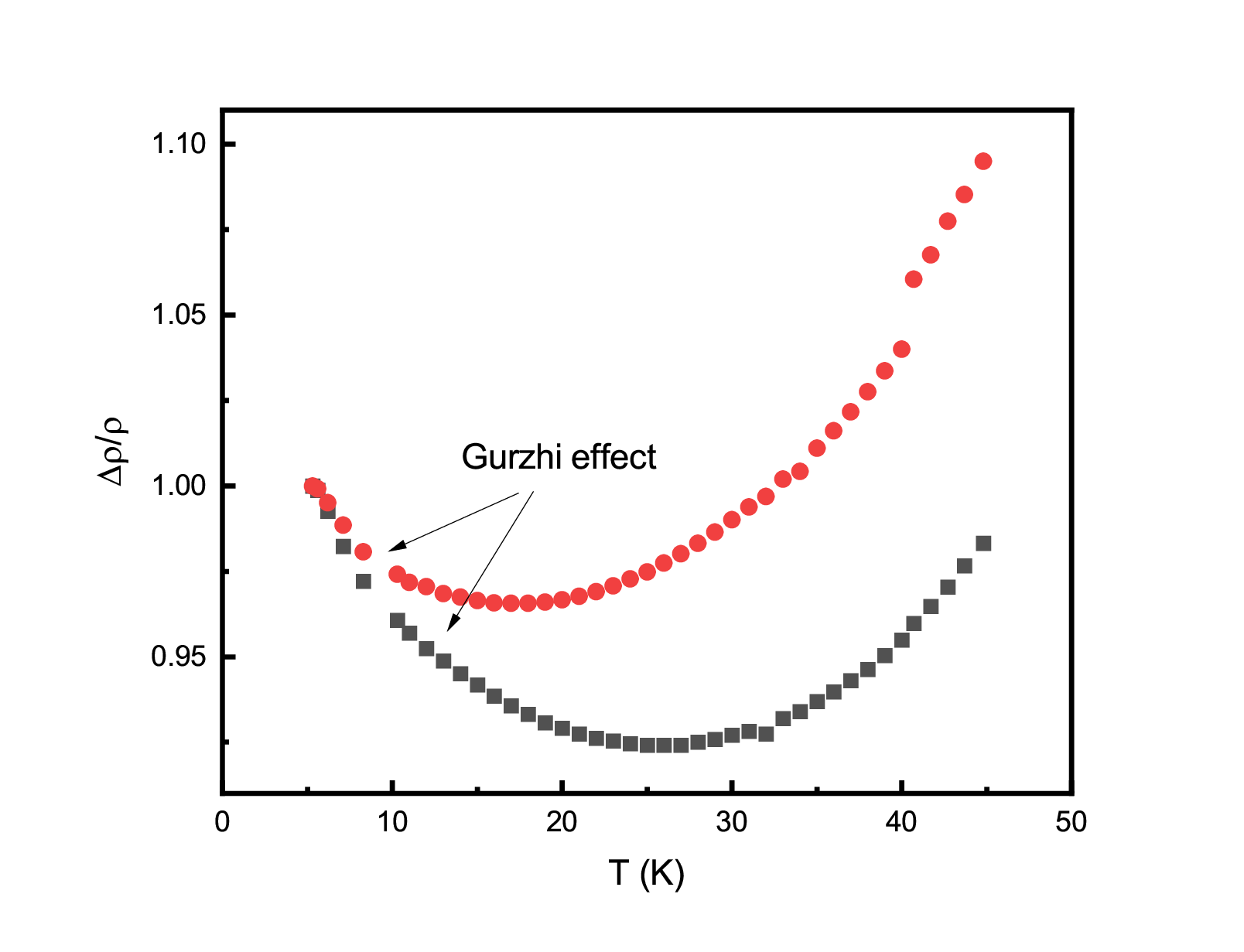}
\caption{ Temperature dependent relative resistivity for 2 triple well, Sample A (red circles), Sample B (black circles).}
\end{figure}
\begin{figure}[ht]
\includegraphics[width=10cm]{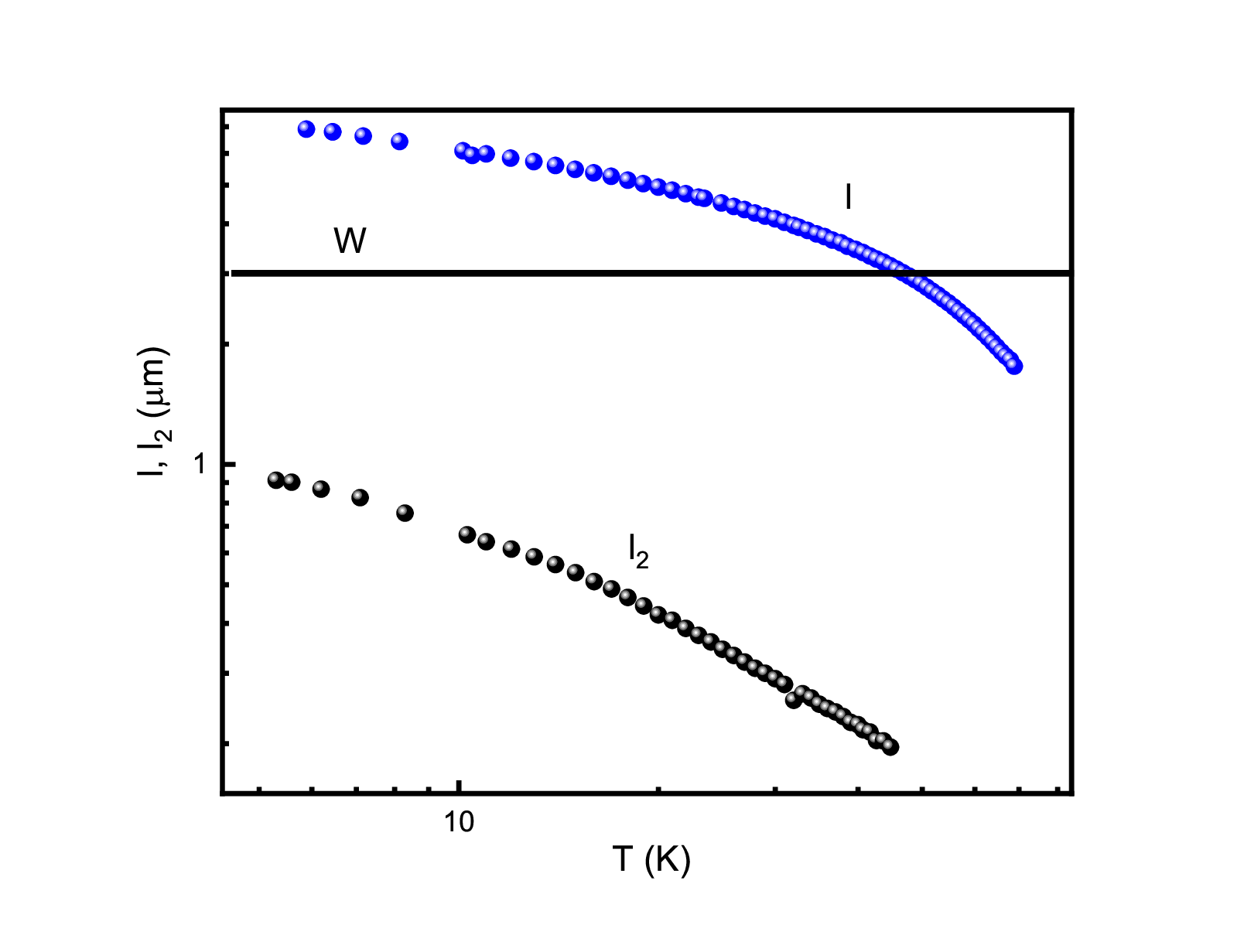}
\caption{ The characteristic lengths $l$ and $l_{2}$ as a function of temperature. The mean free path have been obtained from the measurement in macroscopic sample. Fit of characteristic length with parameters indicated in the main text. Horizontal line - the width of the sample W.}
\end{figure}

The hydrodynamic description is applicable under conditions where  $l_{ee}< W < l $, with $l$ representing the mean free path, $W$ the width of the sample, and $l_{ee}$ the mean free path due to electron-electron collisions. However, it is important to consider the condition $l_2 < l$, where $l_{2} =v _F \tau_{2}$ represents the mean free path associated with the relaxation of the second harmonic of the distribution function \cite{alekseev}. This relaxation likely involves scattering by impurities, affecting the "residual" relaxation rate of shear stress as $T\rightarrow 0$ due to electron scattering on disorder \cite{alekseev}.

In this scenario, the condition $l_{2} < W < l $ is satisfied, indicating that we remain within the hydrodynamic regime even at $T=4.2 K$. Experimentally, this is supported by Figure 6 (see also Figs. 1a and b in the main text), which show that resistance decreases with increasing temperature (the Gurzhi effect), a phenomenon that cannot be explained by ballistic effects alone.  Furthermore, the extracted electron-electron scattering rate, shown in Figure 2 of the main text, follows a $T^2$ dependence, unlike in the ballistic regime where the scattering rate would exhibit a linear T dependence. Additionally, in mesoscopic samples, the scattering rate  $1/\tau$ is typically lower than in macroscopic samples due to boundary scattering and geometric factors.

Figure 7 presents the temperature dependence of the mean free path due to impurity and phonon scattering for macroscopic samples \cite{wiedmann}, while the characteristic length $l_2$ was determined from magnetoresistance measurements for sample B described in the main text. It can be seen that the hydrodynamic condition $l_{2}< W \leq l$ is satisfied across the entire temperature range used in the experiment.

It is also important to note that the condition $l > W $ is quite strict and implies a pure hydrodynamic regime, leading to $\rho = \frac{m} {ne^2} \eta \frac{12} {W^2}$ (Gurzhi regime). However, more recent theories account for an additional term in the resistance equation (1) in the main text, which includes relaxation by phonons or impurities (first harmonic relaxation). Therefore, the condition $W \approx l$ is also relevant for studying viscosity.


\begin{thebibliography}{41}

\bibitem{gurzhi}
R. N. Gurzhi, Minimum of Resistance in Impurity-free Conductors, Sov. Phys. JETP {\bf 44}, 771 (1963); Reviews of Topical Problems: Hydrodynamic Effects in Solids at Low Temperature
Sov. Phys. Usp. {\bf 11}, 255 (1968).

\bibitem{andreev}
A. V. Andreev, S. A. Kivelson, and B. Spivak, Hydrodynamic Description of Transport in Strongly Correlated Electron Systems, Phys. Rev. Lett. {\bf 106}, 256804 (2011).

\bibitem{narozhny2}
B. N. Narozhny, I. V. Gornyi, M. Titov, M. Schutt, and A. D. Mirlin, Hydrodynamics in graphene: Linear-response transport, Phys. Rev. B {\bf 91}, 035414 (2015).

\bibitem{dejong}
M. J. M. de Jong and L. W. Molenkamp, Hydrodynamic electron flow in high-mobility wires, Phys. Rev. B {\bf 51}, 13389 (1995).
\bibitem{gusev1}
G. M. Gusev, A. D. Levin, E. V. Levinson, and A. K. Bakarov, Viscous electron flow in mesoscopic two-dimensional electron gas, AIP Adv. {\bf 8},
025318 (2018).

\bibitem{gusev4}
G. M. Gusev, A. S. Jaroshevich, A. D. Levin, Z. D. Kvon  and A. K. Bakarov, Viscous magnetotransport and Gurzhi effect in bilayer electron system, Phys. Rev. B {\bf 103}, 075303 (2021).


\bibitem{principi}
A. Principi, G. Vignale, M. Carrega, and M. Polini, Bulk and shear viscosities of the two-dimensional electron liquid in a doped graphene sheet, Phys. Rev. B {\bf 93}, 125410 (2016).
\bibitem{alekseev1}
P. S. Alekseev, Negative Magnetoresistance in Viscous Flow of Two-Dimensional Electrons, Phys. Rev. Lett. {\bf 117}, 166601 (2016).
\bibitem{narozhny1}
B. N. Narozhny and M. Schutt, Magnetohydrodynamics in graphene: Shear and Hall viscosities, Phys. Rev. B {\bf 100}, 035125 (2019).
\bibitem{dmitriev}
P. S. Alekseev and A. P. Dmitriev, Viscosity of two-dimensional electrons, Phys. Rev. B {\bf 102}, 241409(R) (2020).
\bibitem{raichev2}
O. E. Raichev, G. M. Gusev, A. D. Levin, and A. K. Bakarov, Manifestations of classical size effect and electronic viscosity in the magnetoresistance of narrow two-dimensional conductors: Theory and experiment, Phys. Rev. B {\bf 101}, 235314 (2020).
\bibitem{gusev5}
D. A. Khudaiberdiev,  G. M. Gusev,  E. B. Olshanetsky, Z. D. Kvon,  and N. N. Mikhailov, Magnetohydrodynamics and electron-electron interaction of massless Dirac fermions,
Phys. Rev. Research {\bf 3}, L032031 (2021).
\bibitem{du}
Xinghao Wang, Peizhe Jia, Rui-Rui Du, L. N. Pfeiffer, K. W. Baldwin, and K. W. West,
Hydrodynamic charge transport in an GaAs/AlGaAs ultrahigh-mobility two-dimensional electron gas, Phys. Rev. B {\bf 106}, L241302 (2022).
\bibitem{gusev6}
A. D. Levin, G. M. Gusev, A. S. Yaroshevich, Z. D. Kvon, and A. K. Bakarov,  Geometric engineering of viscous magnetotransport in a two-dimensional electron system, Phys. Rev. B  {\bf 108}, 115310 (2023).

\bibitem{bandurin1}
D. A. Bandurin, I. Torre, R. Krishna Kumar, M. Ben Shalom,
A. Tomadin, A. Principi, G. H. Auton, E. Khestanova, K. S.
Novoselov, I. V. Grigorieva, L. A. Ponomarenko, A. K. Geim,
and M. Polini, Negative local resistance caused by viscous electron backflow in graphene, Science {\bf 351}, 1055 (2016).

\bibitem{torre}
I. Torre, A. Tomadin, A. K. Geim, and M. Polini, Nonlocal transport and the hydrodynamic shear viscosity in graphene, Phys. Rev. B {\bf 92}, 165433 (2015).

\bibitem{pellegrino2}
F. M. D. Pellegrino, I. Torre, and M. Polini, Nonlocal transport and the Hall viscosity of two-dimensional hydrodynamic electron liquids, Phys.Rev. B {\bf 96}, 195401 (2017).


\bibitem{levin}
A. D. Levin, G. M. Gusev, E. V. Levinson, Z. D. Kvon, and A. K. Bakarov, Vorticity-induced negative nonlocal resistance in a viscous two-dimensional electron system,
Phys. Rev. B {\bf 97}, 245308 (2018).
\bibitem{kumar}
R. Krishna Kumar, D. A. Bandurin, F. M. D.Pellegrino, Y. Cao, A. Principi, H. Guo, G. H. Auton, M. Ben Shalom, L. A. Ponomarenko, G. Falkovich, K. Watanabe, T. Taniguchi, I. V. Grigorieva, L. S.Levitov, M.Polini, and A. K.Geim, Superballistic flow of viscous electron fluid through graphene constrictions, Nat. Phys. {\bf 13}, 1182 (2017).

\bibitem{holder}
T. Holder, R. Queiroz, T. Scaffidi, N. Silberstein, A. Rozen, J. A. Sulpizio,
L. Ella, S. Ilani, and A. Stern, Ballistic and hydrodynamic magnetotransport in narrow channels, Phys. Rev. B {\bf 100}, 245305 (2019).
\bibitem{lucas}
A. Lucas, Stokes paradox in electronic Fermi liquids. Phys. Rev. B {\bf 95}, 115425 (2017).
\bibitem{gusev7}
G. M. Gusev, A. S. Yaroshevich, A.D.Levin, Z. D. Kvon, A.K.Bakarov, Stokes flow around an obstacle in viscous two-dimensional electron liquid, Sci.Rep., {\bf 10}, 7860 (2020).
\bibitem{gornyi}
I. V. Gornyi and D. G. Polyakov, Two-dimensional electron hydrodynamics in a random array of impenetrable obstacles: Magnetoresistivity, Hall viscosity, and the Landauer dipole, Phys.Rev. B 108, 165429 (2023).
\bibitem{krebs}
Zachary J. Krebs, Wyatt A. Behn, Songci Li, Keenan J. Smith, Kenji Watanabe, Takashi Taniguchi, Alex Levchenko, Victor W. Brar, Imaging the breaking of electrostatic dams in
graphene for ballistic and viscous fluids, Science {\bf 379}, 671 (2023).
\bibitem{pusep}
Yu. A. Pusep,  M. D. Teodoro, V. Laurindo, Jr., E. R. Cardozo de Oliveira , G. M. Gusev , and A. K. Bakarov, Diffusion of photoexcited holes in a viscous electron fluid, Phys. Rev. Lett., {\bf 128}, 136801 (2022).
\bibitem{pusep2}
M. A. T. Patricio, G. M. Jacobsen, M. D. Teodoro, G. M. Gusev,
A. K. Bakarov, and Yu. A. Pusep, Hydrodynamics of electron-hole fluid photogenerated in a mesoscopic two-dimensional channel, Phys. Rev. B {\bf 109}, L121401 (2024).

\bibitem{berdyugin}
A. I. Berdyugin, S. G. Xu, F. M. D. Pellegrino, R. Krishna
Kumar, A. Principi, I. Torre, M. Ben Shalom, T. Taniguchi,
K. Watanabe, I. V. Grigorieva, M. Polini, A. K. Geim, and
D. A. Bandurin, Measuring Hall viscosity of graphene{\textquoteright}s electron fluid, Science {\bf 364}, 162 (2019).
\bibitem{scaffidi}
T. Scaffidi, N. Nandi, B. Schmidt, A. P. Mackenzie, and J. E. Moore, Hydrodynamic Electron Flow and Hall Viscosity, Phys. Rev. Lett. {\bf 118}, 226601 (2017).

\bibitem{burmistrov}
I. S. Burmistrov, M. Goldstein, M. Kot, V. D. Kurilovich, and
P. D. Kurilovich, Dissipative and Hall Viscosity of a Disordered 2D Electron Gas, Phys. Rev. Lett. {\bf 123}, 026804 (2019).

\bibitem{alekseev2}
P. S. Alekseev and M. A. Semina, Ballistic flow of two-dimensional interacting electrons, Phys. Rev. B {\bf 98}, 165412 (2018).

\bibitem{alekseev4}
P. S. Alekseev and M. A. Semina, Hall effect in a ballistic flow of two-dimensional interacting particles, Phys. Rev. B {\bf 100}, 125419 (2019).

\bibitem{gusev2}
G. M. Gusev, A. D. Levin, E. V. Levinson, and A. K. Bakarov, Viscous transport and Hall viscosity in a two-dimensional electron system, Phys. Rev. B {\bf 98}, 161303(R) (2018).


\bibitem{polini}
Marco Polini, and Andre K. Geim, Viscous electron fluids, Physics Today {\bf 73}, 6, 28 (2020).


\bibitem{narozhny}
Boris N. Narozhny, Hydrodynamic approach to two-dimensional electron
systems, La Rivista del Nuovo Cimento, {\bf 45}, 661–736 (2022).
\bibitem{bradlyn}
B. Bradlyn, M. Goldstein, and N. Read, Kubo formulas for viscosity: Hall viscosity, Ward identities, and the relation with conductivity, Phys. Rev. B {\bf 86},
245309 (2012).

\bibitem{burmistrov2}
V. A. Zakharov and I. S. Burmistrov, Residual bulk viscosity of a disordered two-dimensional electron gas,  Phys. Rev. B {\bf 103}, 235305  (2021).
\bibitem{sharma}
Bhanuday Sharma, Rakesh Kumar, A brief introduction to bulk viscosity of fluids, physics arXiv:2303.08400, (2023).
\bibitem{khalatnikov}
I. Khalatnikov, The second viscosity of monoatomic gases, Sov. Phys. JETP {\bf 2}, 169 (1955).
\bibitem{cramer}
M. S. Cramer, Numerical estimates for the bulk viscosity of ideal gases.", Phys. Fluids, {\bf 24}, 066102 (2012).
\bibitem{alekseev5}
P. S. Alekseev, Viscous Flow of Two-Component Electron Fluid in Magnetic Field, Semiconductors, 2023, {\bf 57},  193 ( 2023).
\bibitem{alekseev6}
P. S. Alekseev, A. P. Dmitriev, I. V. Gornyi, V. Yu. Kachorovskii, B. N. Narozhny, M. Schutt, M. Titov. Magnetoresistance in Two-Component Systems, Phys. Rev. Lett., {\bf 114}, 156601 (2015).
\bibitem{alekseev7}
P. S. Alekseev, A. P. Dmitriev, I. V. Gornyi, V. Yu. Kachorovskii, B. N. Narozhny, M. Schutt, M. Titov, Magnetoresistance of compensated semimetals in confined geometries, Phys. Rev. B, {\bf 95}, 165410 (2017).

\bibitem{wiedmann}
S. Wiedmann, N. C. Mamani, G. M. Gusev, O. E. Raichev, A. K. Bakarov, J. C.Portal, Magnetoresistance oscillations in multilayer systems: Triple wells, Phys. Rev. B, {\bf 80}, 245306 (2009).
\bibitem{wiedmann2}
S. Wiedmann, N. C. Mamani, G. M. Gusev, O. E. Raichev, A. K. Bakarov, J. C. Portal, Magneto-intersubband oscillations in triple quantum wells, Physica E,  {\bf 42}, 1088 (2010).
\bibitem{wiedmann3}
G. M. Gusev, S. Wiedmann, O. E. Raichev, A. K. Bakarov, J. C. Portal, Emergent and reentrant fractional quantum Hall effect in trilayer systems in a tilted magnetic field, Phys. Rev. B, {\bf 80}, 161302(R) (2009).
 \bibitem{suppl}
 See Supplemental Material at http://link.aps.org/ for details of the triple quantum well and transport analyses. The Supplemental Material also contains Refs. [10, 43-45].
\bibitem{slutzky}
M. Slutzky, O. Entin-Wohlman, Y. Berk, A. Palevski, and H. Shtrikman, Electron-electron scattering in coupled quantum wells, Phys. Rev. B {\bf 53}, 4065 (1996).
\bibitem{holten}
M. Holten, L. Bayha, A. C. Klein, P. A. Murthy, P. M. Preiss, and
S. Jochim, Anomalous breaking of scale invariance in a two-dimensional Fermi gas, Phys. Rev. Lett. {\bf 121}, 120401 (2018).
\end{thebibliography}

\begin{thebibliography}{41}

\bibitem{wiedmann}
S. Wiedmann, N. C. Mamani, G. M. Gusev, O. E. Raichev, A. K. Bakarov, J. C.Portal, Magnetoresistance oscillations in multilayer systems: Triple wells, Phys. Rev. B, 2009, {\bf 80}, 245306 (2009).
\bibitem{wiedmann2}
S. Wiedmann, N. C. Mamani, G. M. Gusev, O. E. Raichev, A. K. Bakarov, J. C. Portal, Magneto-intersubband oscillations in triple quantum wells, Physica E,  {\bf 42}, 1088 (2010).
\bibitem{wiedmann3}
G. M. Gusev, S. Wiedmann, O. E. Raichev, A. K. Bakarov, J. C. Portal, Emergent and reentrant fractional quantum Hall effect in trilayer systems in a tilted magnetic field, Phys. Rev. B, {\bf 80}, 161302(R) (2009).
\bibitem{alekseev}
P. S. Alekseev and A. P. Dmitriev, Viscosity of two-dimensional electrons, Phys. Rev. B {\bf 102}, 241409(R) (2020).
\end{thebibliography}
\end{document}